\documentclass[10pt,final,journal,oneside,twocolumn]{CustomIEEE_CopyRight}
\usepackage[pdftex]{graphicx}
\usepackage{color}
\usepackage{transparent}
\usepackage{tabularx}
\usepackage{makecell}
\usepackage{subfig}
\usepackage{graphicx}
\usepackage[cmex10]{amsmath}
\usepackage[noadjust]{cite}
\usepackage[font=normal,labelfont = bf,textfont=it]{caption}
\renewcommand{\refname}{\textbf{\textit{\textsl{References}}}} 

\newcolumntype{L}[1]{>{\raggedright\let\newline\\\arraybackslash\hspace{0pt}}m{#1}}
\newcolumntype{C}[1]{>{\centering\let\newline\\\arraybackslash\hspace{0pt}}m{#1}}
\newcolumntype{R}[1]{>{\raggedleft\let\newline\\\arraybackslash\hspace{0pt}}m{#1}}

\begin{document}
\title{On the Influence of Spatial Dispersion on the Performance of Graphene-Based Plasmonic Devices}

\author{
\quad \quad\quad\quad Diego Correas-Serrano,
Juan Sebastian~Gomez-Diaz,~\IEEEmembership{Member,~IEEE}
and \newline Alejandro~Alvarez-Melcon,~\IEEEmembership{Senior Member,~IEEE}

\thanks{D. Correas-Serrano and A. Alvarez-Melcon are with the Departamento de Tecnologias de la Informacion y las Comunicaciones, Universidad Politecnica de Cartagena, Cartagena E-30202 (Murcia), Spain (e-mail: diego.correas.serrano@gmail.com, alejandro.alvarez@upct.es)}

\thanks{%
J. S. Gomez-Diaz is with the Adaptive MicroNanoWave Systems, LEMA/Nanolab,
\'Ecole Polytechnique F\'ed\'erale de Lausanne.
1015 Lausanne, Switzerland
(e-mail: juan-sebastian.gomez@epfl.ch)
}}
\maketitle

\begin{abstract}
We investigate the effect of spatial dispersion phenomenon on the performance of graphene-based plasmonic devices at THz. For this purpose, two different components, namely a phase shifter and a low-pass filter, are taken from the literature, implemented in different graphene-based host waveguides, and analyzed as a function of the surrounding media. In the analysis, graphene conductivity is modeled first using the Kubo formalism and then employing a full-$k_\rho$ model which accurately takes into account spatial dispersion. Our study demonstrates that spatial dispersion up-shifts the frequency response of the devices, limits
their maximum tunable range, and degrades their frequency response. Importantly, the influence of this phenomenon significantly increases with higher permittivity values of the surrounding media, which is related to the large  impact of spatial dispersion in very slow waves.
These results confirm the necessity of accurately assessing non-local effects in the development of practical plasmonic THz devices.
\end{abstract}


\IEEEpeerreviewmaketitle
\section{\textit{\textsf{\textbf{Introduction}}}}
Graphene has recently attracted large attention as a potential platform for surface plasmon polaritons (SPPs) at terahertz (THz) frequencies \cite{Koppens11}. Graphene electrical properties, which include high carrier mobility and saturation speed, electrostatic and magnetostatic reconfiguration, and relatively low losses among many others \cite{Geim2007}, have triggered the theoretical development of tunable plasmonic devices such as waveguides \cite{Christensen12},
leaky-wave antennas \cite{Sebas13_LWA_width_modulated, marc2013leaky}, reflectarrays \cite{Carrasco13_AWPL}, switches \cite{sebas13_Switch}, phase shifters \cite{Chen13_phase_shifters}, or filters \cite{Diego13}. Importantly, the propagation of SPPs along graphene has just been experimentally confirmed \cite{Chen12} and some initial devices, including THz modulators \cite{Sensale-Rodriguez12} and Faraday rotators \cite{Sounas13_Faraday}, have been demonstrated. Consequently, it is expected that novel plasmonic devices with unprecedent performance in the THz band will be developed in the coming years.

\begin{figure}[!t] \centering
\subfloat[]{\label{fig:_MainFiga}
\includegraphics[width=0.8\columnwidth]{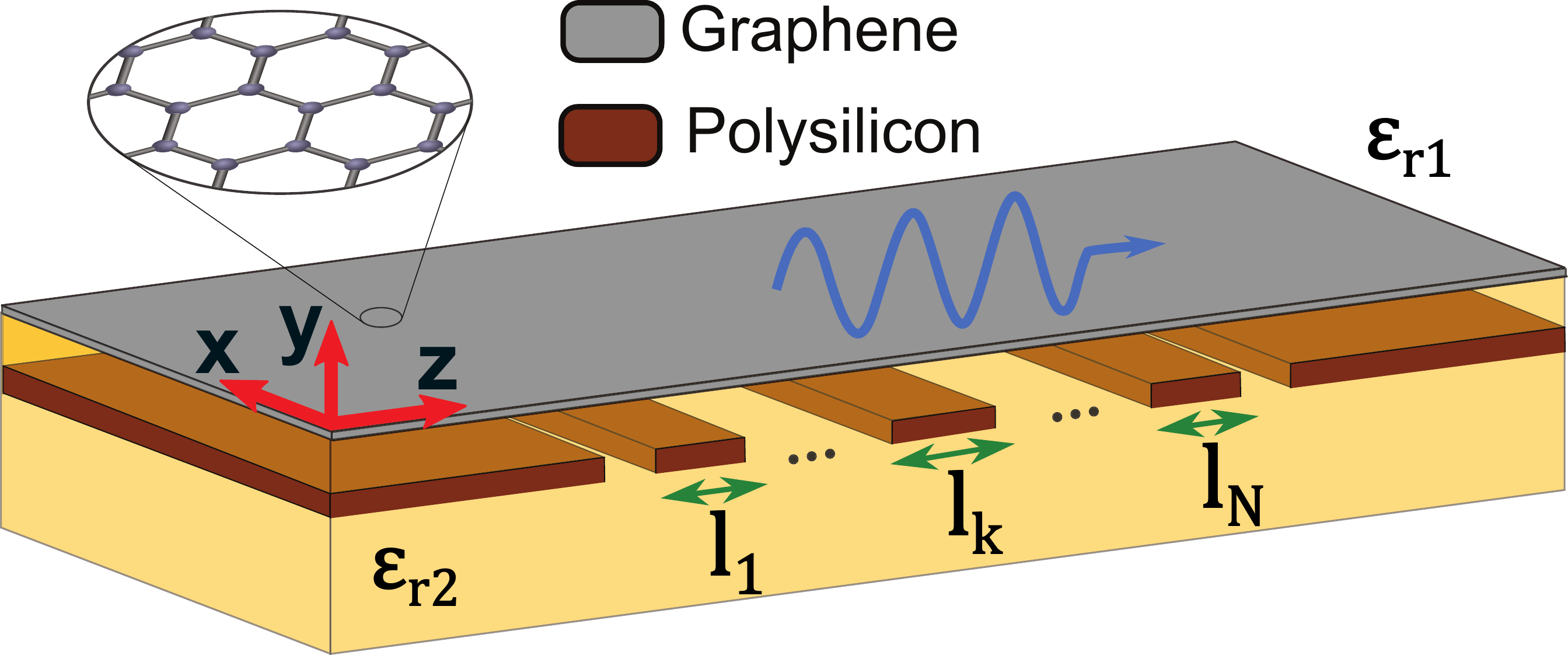}}\\
\subfloat[]{\label{fig:_MainFigb}
\includegraphics[width=0.8\columnwidth]{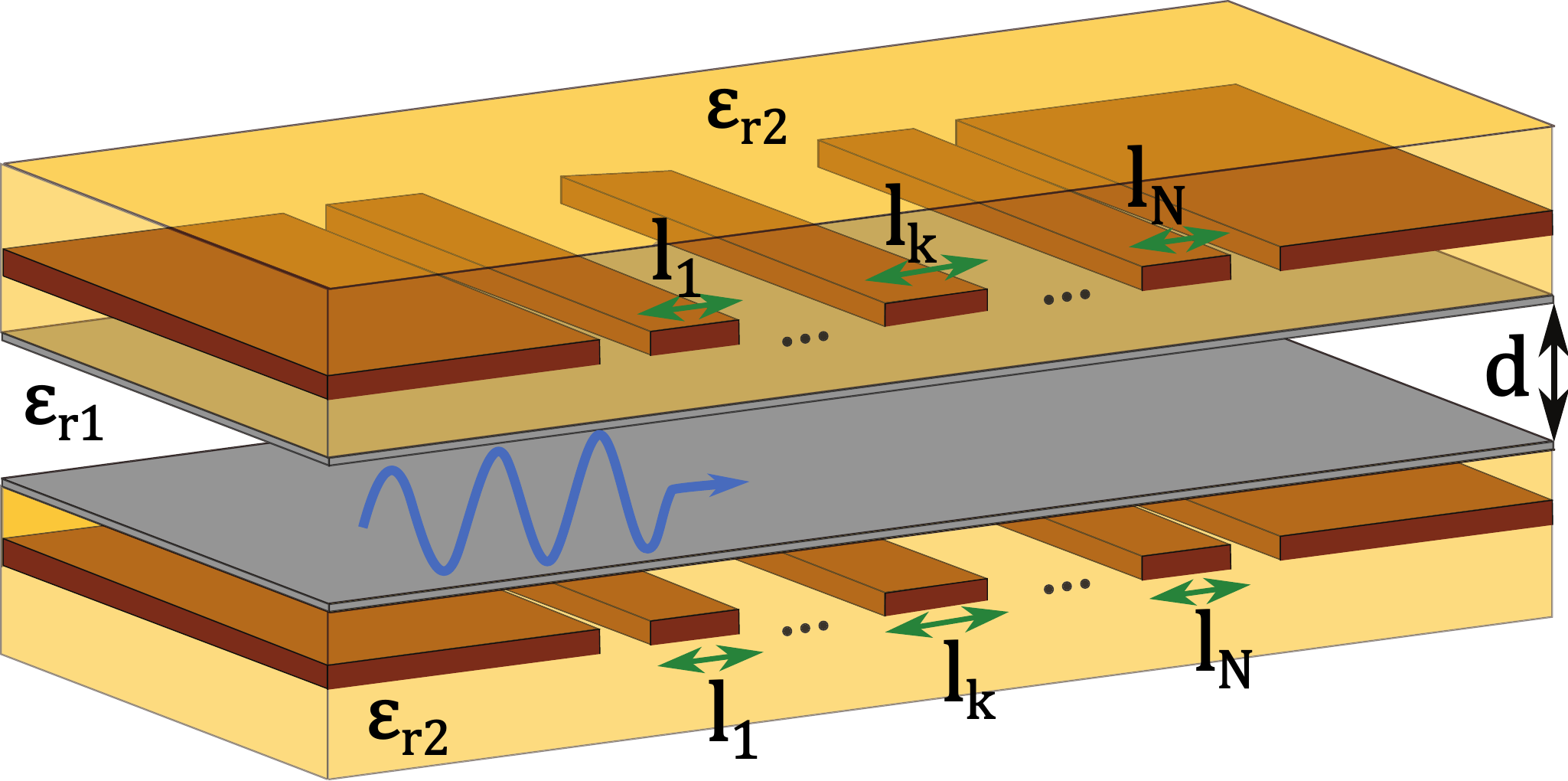}}\\
\caption{Schematic diagram of two graphene-based waveguides. Graphene conductivity may be tuned at each waveguide section by applying a DC bias to the poliysilicon gating pads. (a) Monolayer implementation. (b) PPW implementation.} \label{Main_draw}
\end{figure}

In the theoretical development of non-magnetic plasmonic devices, graphene has usually been modeled as an infinitesimally thin layer characterized by a scalar conductivity obtained through the Kubo formalism \cite{Hanson09}. This formulation provides a large scale model of graphene, valid from DC to optical frequencies, but neglects the possible presence of spatial dispersion (non-local) effects, i.e. the dependence of graphene conductivity with the propagating wavenumber. Several authors have recently investigated how this phenomenon modifies the propagation characteristics of SPPs \cite{Hanson09, lovat2013waveguides, Sebas13_Spatial_dispersion, Hanson13, Diego13_MTT}, concluding that spatial dispersion
becomes
significant in case of very large propagating wavenumbers.

In this context, this letter studies the influence of spatial dispersion effects
on the performance of graphene-based plasmonic devices at THz. For this purpose, two examples of such devices, namely a phase shifter \cite{Chen13_phase_shifters} and a low-pass filter \cite{Diego13}, have been chosen from the literature, implemented in graphene-based single layer and parallel-plate (PPW) host waveguides \cite{lovat2013waveguides, Diego13_MTT}, and analyzed as a function of the surrounding media. In the analysis, graphene is modeled first using the Kubo formalism
\cite{Hanson09} and then employing an accurate conductivity model which
takes into account spatial dispersion in the THz band \cite{Hanson13}. Our study demonstrates that (a) spatial dispersion modifies the behavior of plasmonic devices by up-shifting their operation frequency, degrading their frequency response, and limiting their
maximum tunable range; and (b) the influence of spatial dispersion increases with higher permittivities of the surrounding media. These results clearly confirm the necessity of accurately assessing spatial dispersion phenomenon in the development of plasmonic THz devices.
\section{\textit{\textsf{\textbf{Graphene-based Host Waveguides}}}}
This section briefly describes the behavior of two well-known $2$D graphene-based waveguides \cite{Hanson09, Chen13_phase_shifters} (see Fig.~\ref{Main_draw}) which are employed below to design the plasmonic devices under investigation.
The lateral dimensions of the waveguides are assumed to be much larger than the guided wavelength.
The first configuration (see Fig.~\ref{fig:_MainFiga}) consists of a single graphene layer transferred onto a dielectric, and it supports a unique TM SPP \cite{Hanson09} at the frequency band of interest. The second structure (see Fig.~\ref{fig:_MainFigb}) is composed of a
graphene-based parallel-plate waveguide
with a separation distance $d$ between the graphene sheets much smaller than their width. Even though this waveguide may support two modes \cite{Christensen12, lovat2013waveguides}, we focus
here on the
quasi-TEM mode due to its extreme confinement to the graphene layers. In both waveguides, gating pads are located close to the graphene layers to dynamically control the guiding properties of each waveguide section
by exploiting
graphene's field effect \cite{Geim2007, Chen13_phase_shifters}.

\begin{figure}\label{fig: cmaps} \centering
\subfloat[]{\label{fig: colormapa}
\includegraphics[width=0.5\columnwidth]{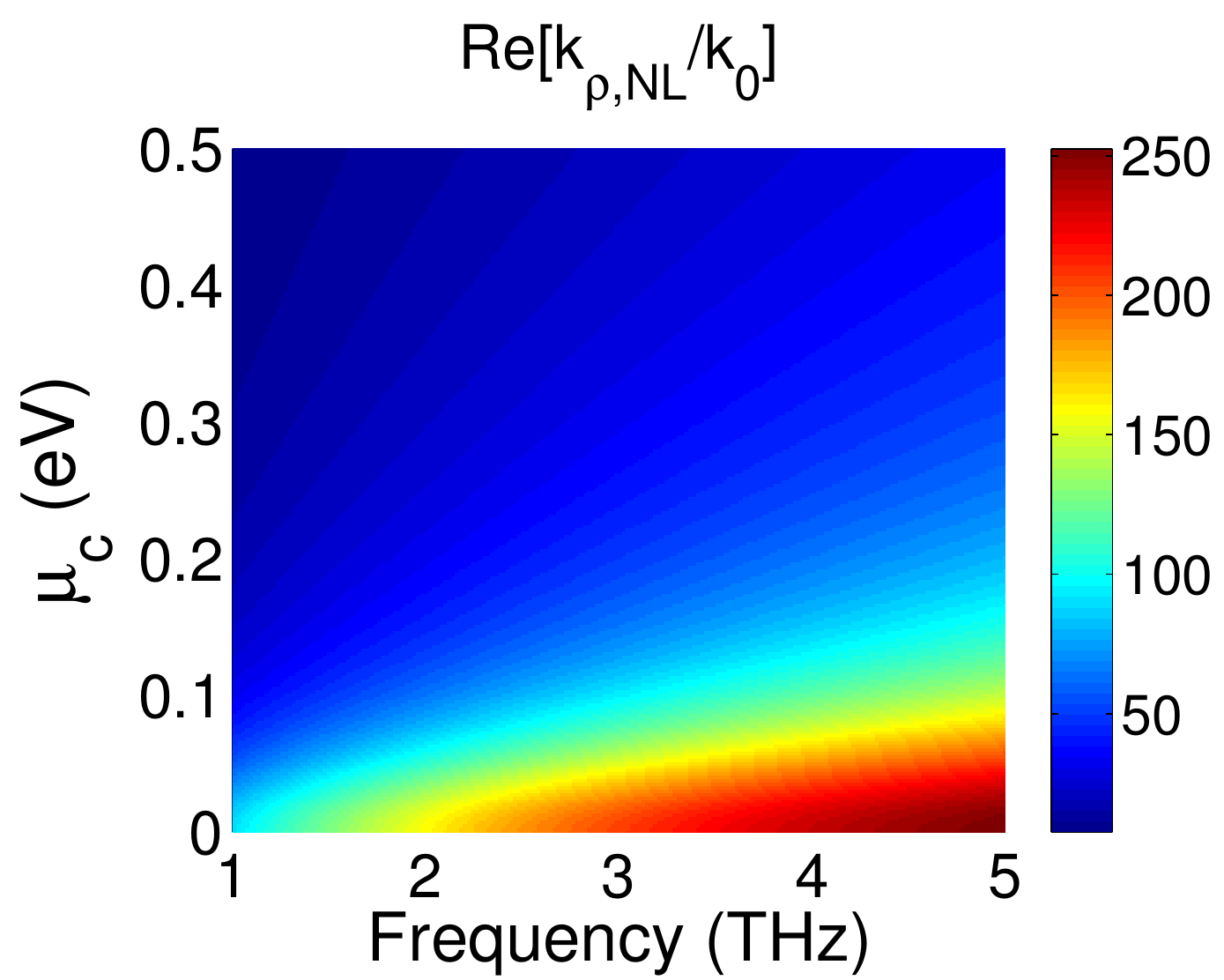}}
\subfloat[]{\label{fig: colormapb}
\includegraphics[width=0.5\columnwidth]{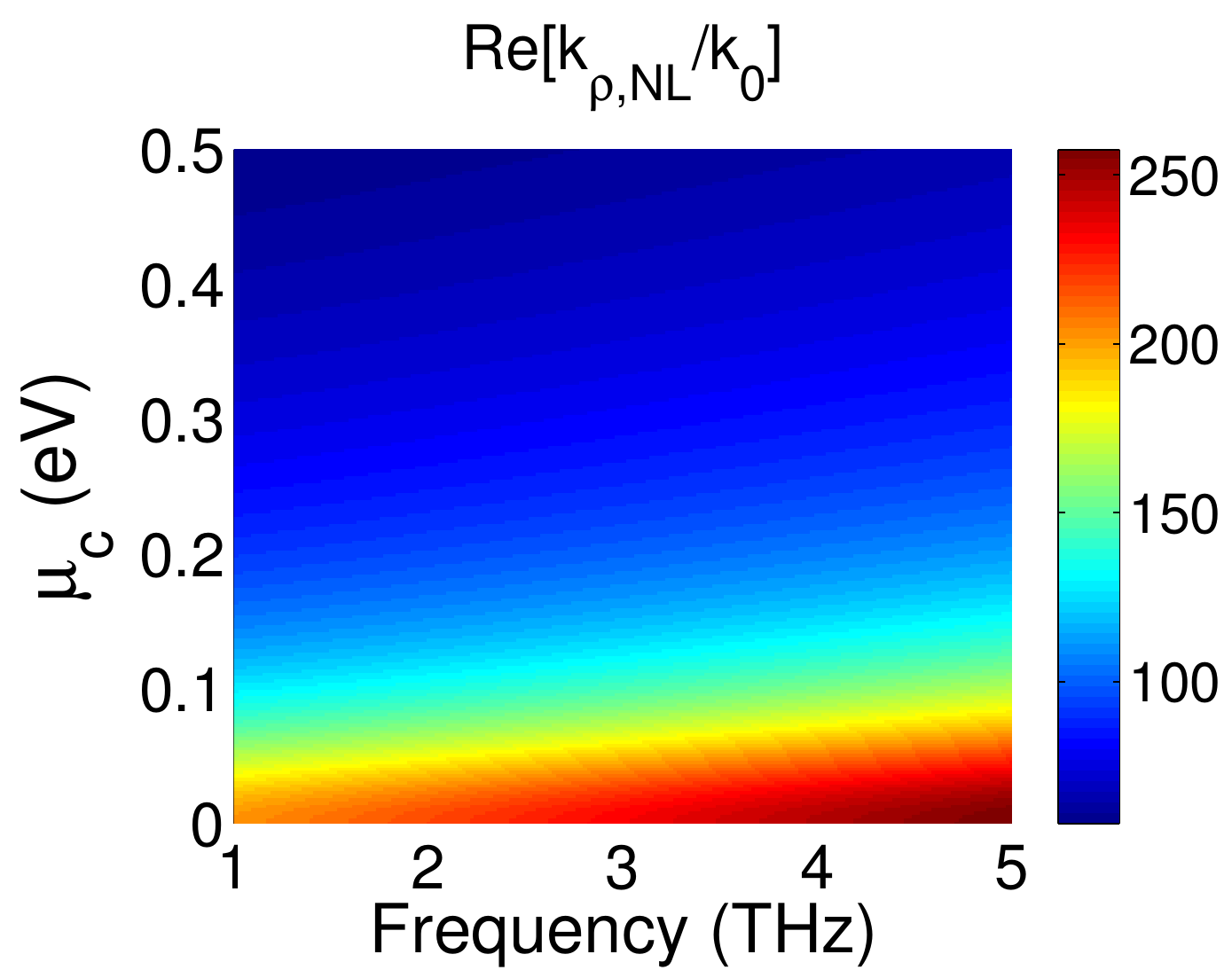}}\\
\subfloat[]{\label{fig: colormapc}
\includegraphics[width=0.5\columnwidth]{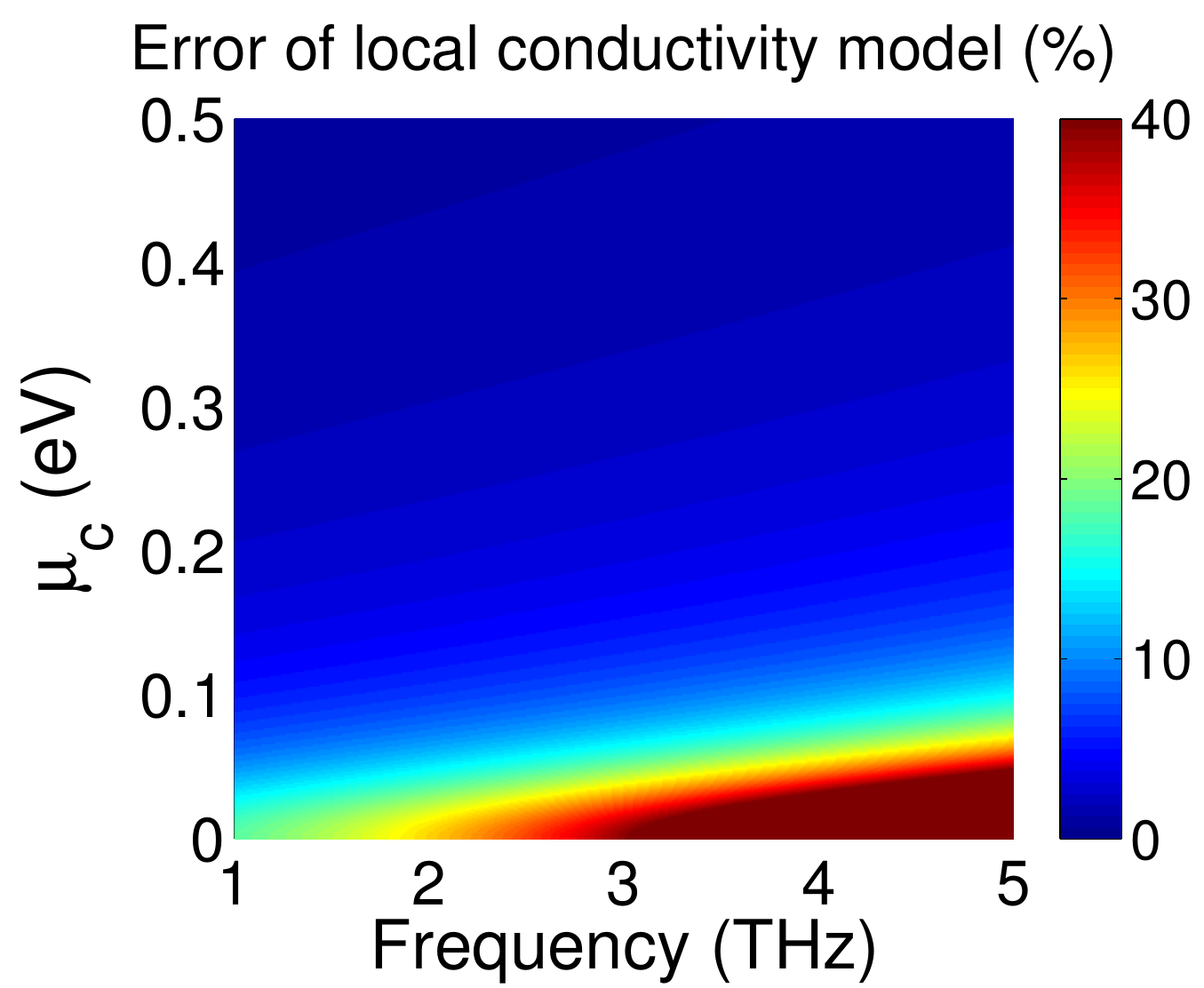}}
\subfloat[]{\label{fig: colormapd}
\includegraphics[width=0.5\columnwidth]{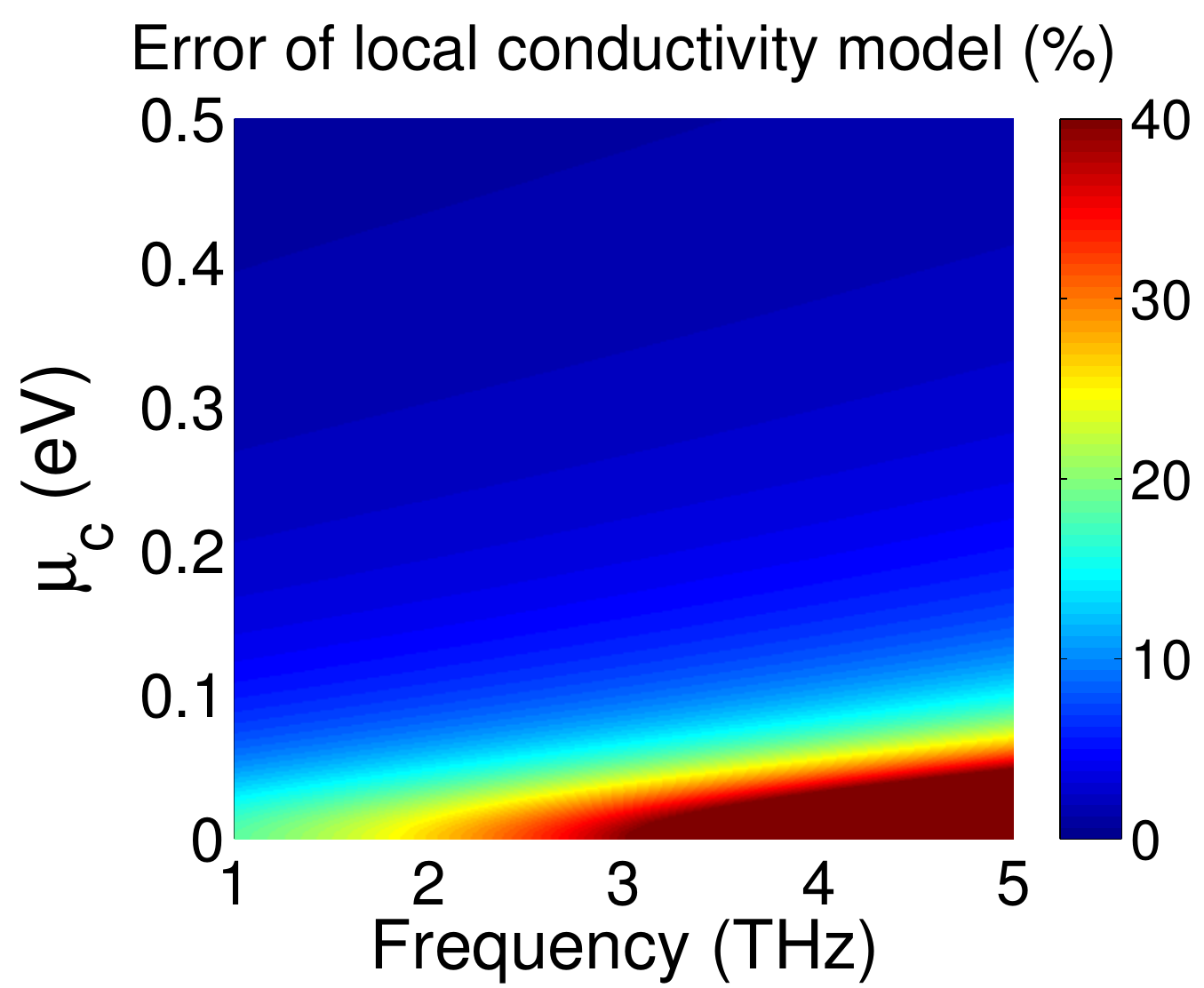}}
\caption{Normalized phase constant of (a) TM mode in a monolayer graphene waveguide (see Fig. \ref{fig:_MainFiga}) and (b) quasi-TEM mode in a graphene-based PPW (see Fig. \ref{fig:_MainFigb}). (c)-(d) show the relative error $100\left|\frac{k_{\rho_L}-k_{\rho_{NL}}}{k_{\rho_{NL}}}\right|$ in the calculation of panels (a)-(b) when using a local conductivity model. Parameters are $d = 100$~nm, $\tau = 1$~ps, $\varepsilon_{r1} = \varepsilon_{r2} = 11.9$ and $T=300$~K.}\label{fig: colormaps}
\end{figure}

In the analysis of the host waveguides, graphene is modeled as an infinitesimally thin layer characterized by a scalar conductivity which depends on frequency, graphene's electron relaxation time $\tau$ and chemical potential $\mu_c$, and temperature $T$. Neglecting the influence of spatial dispersion, graphene's conductivity is obtained using the well-known Kubo formalism \cite{Hanson09}.
Then, the propagation constant $k_{\rho_L}$ and characteristic impedance $Z_{C_L}$
of the propagating modes can be easily obtained using standard techniques \cite{Hanson09, Chen13_phase_shifters}. When spatial dispersion effects are taken into account, graphene is characterized using the full-$k_\rho$
relaxation-time-approximation (RTA)
conductivity model introduced in \cite{Hanson13}, which provides
accurate
representation for any plasmon wavenumber and is valid in the whole THz band.
Importantly, we have numerically verified that the Bhatnagar-Gross-Krook conductivity model also proposed in \cite{Hanson13}, which enforces charge conservation and is expected to be more accurate, leads to extremely similar results \cite{Diego13_MTT} while adding substantial mathematical complexity.
The propagation constant $k_{\rho_{NL}}$ and characteristic impedance $Z_{C_{NL}}$ of the host waveguides can then be obtained in closed-form following the approach developed in \cite{Diego13_MTT}.
For the sake of illustration, Figs.~\ref{fig: colormapa}-\ref{fig: colormapb} show the normalized phase constant of the considered spatially-dispersive
graphene waveguides embedded in silicon (Si) ($\varepsilon_{r} = 11.9$)
as a function of frequency and chemical potential. Figs. \ref{fig: colormapc}-\ref{fig: colormapd} show the relative error in these computations when neglecting spatial dispersion. As can be observed, the Kubo conductivity model systematically overestimates the
propagation constants of the waves, especially when graphene chemical potential is low. In addition, the non-linear distribution of this error as a function of $\mu_c$ will inevitably lead to unexpected behavior of graphene-based plasmonic devices. Very large
wavenumbers which were possible to obtain in the local case
using low chemical potentials, can no longer be obtained due
to spatial dispersion.
This can be understood in terms of the finite quantum capacitance of graphene \cite{Hanson13, Diego13}, which impairs large values of the wavevector. This limitation implies that plasmonic devices that rely on abrupt variations of graphene conductivity may become more difficult to realize in practice than initially expected.
%
%
\section{\textit{\textsf{\textbf{Results and discussion}}}}
This section studies the influence of spatial dispersion on the performance of graphene-based plasmonic phase shifters \cite{Chen13_phase_shifters} and low-pass filters \cite{Diego13}. Results have been obtained using a transmission line model combined with a transfer-matrix approach \cite{Chen13_phase_shifters, Diego13}, considering graphene with $\tau=1.0$~ps at T=$300$~K.
Without loss of generality, we assume $\varepsilon_{r1} = \varepsilon_{r2} = \varepsilon_r$ throughout this section.
 In the local case, results have been validated using data from full-wave simulations (not shown here for the sake of compactness). However, note that current commercial full-wave software packages are not able to simulate spatially-dispersive graphene structures.
\subsection{\textit{\textsf{Graphene-Based Phase Shifters}}}
\begin{figure*}\centering
\subfloat[]{\label{fig: shiftera}
\includegraphics[width=0.6\columnwidth]{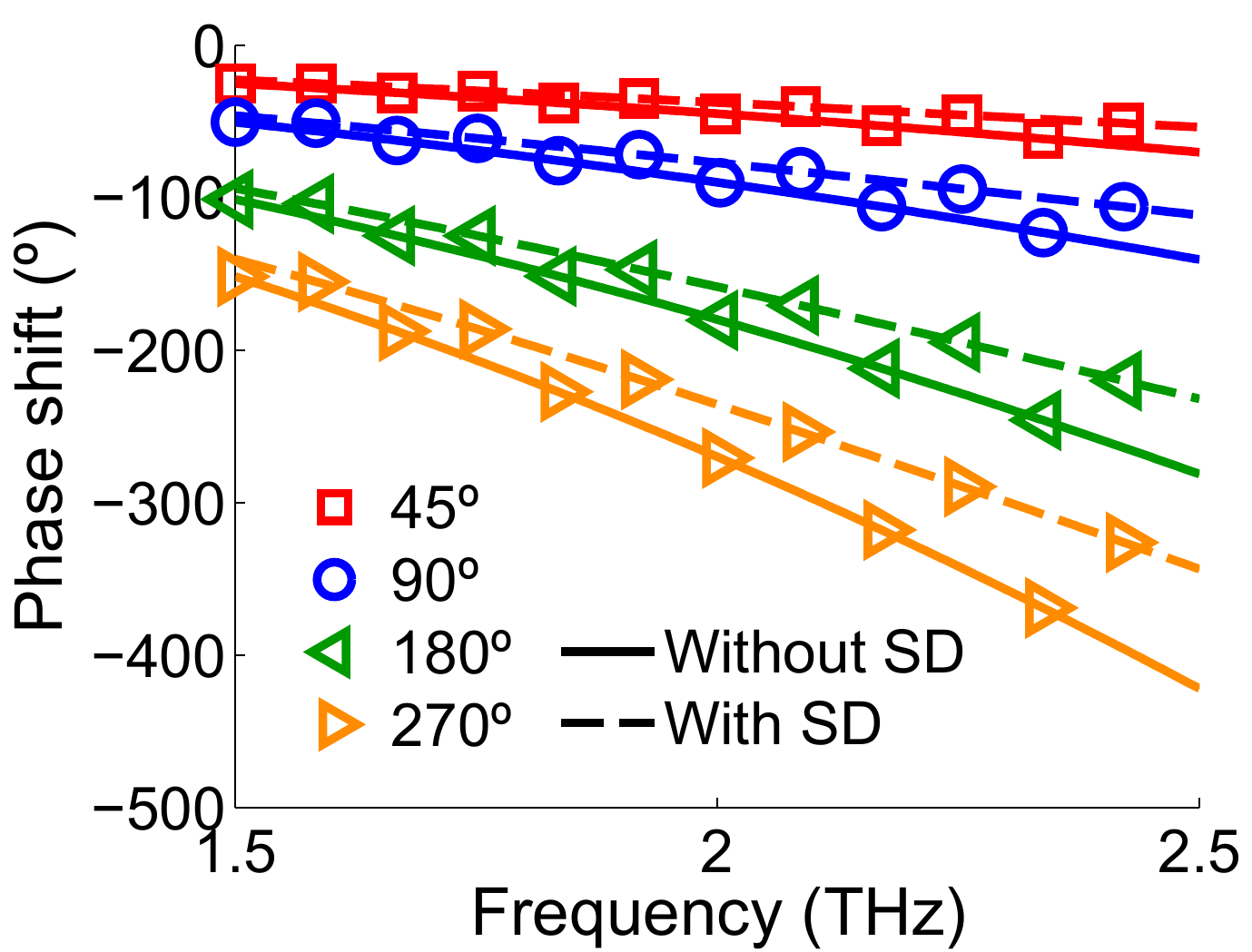}}
\subfloat[]{\label{fig: shifterb}
\includegraphics[width=0.6\columnwidth]{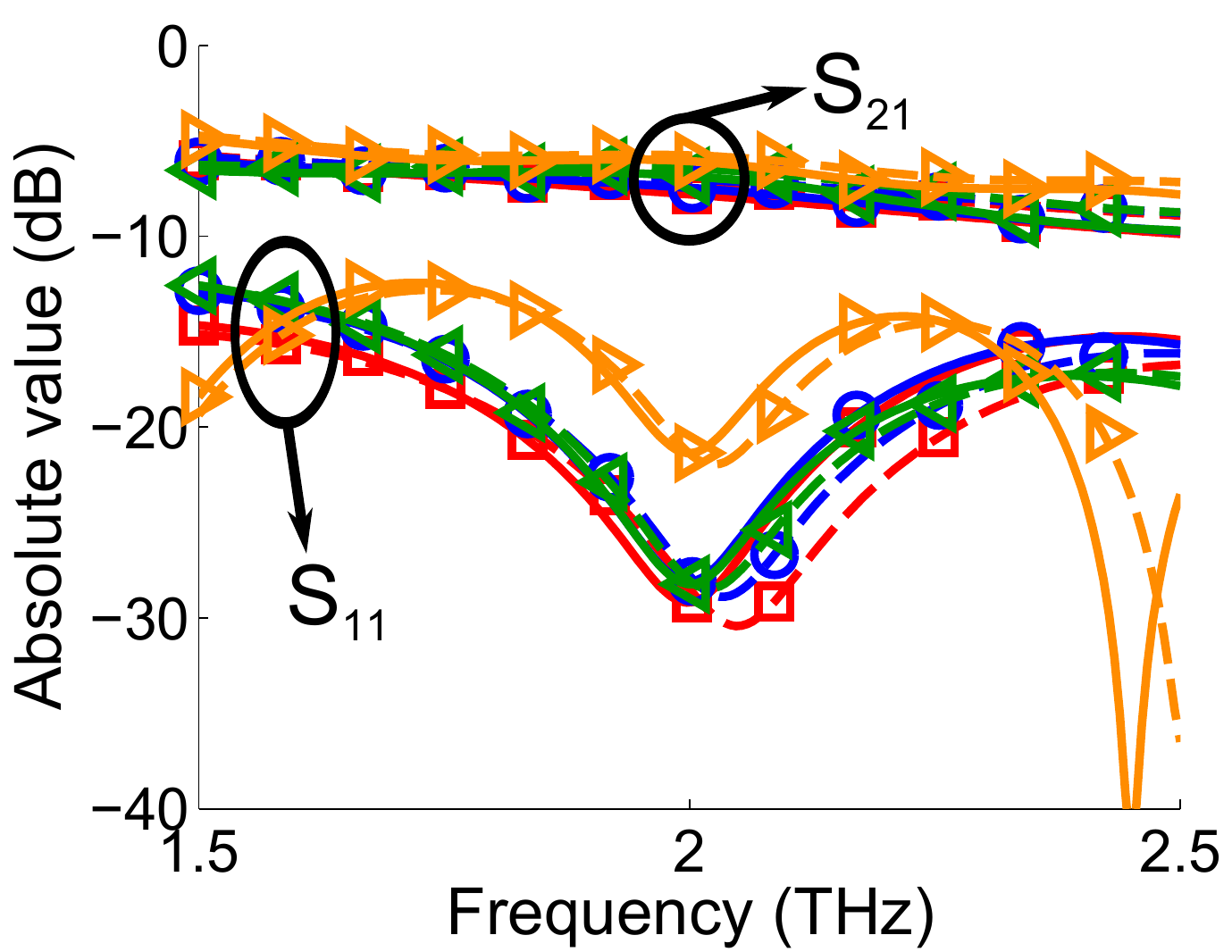}}
\subfloat[]{\label{fig: shifterc}
\includegraphics[width=0.6\columnwidth]{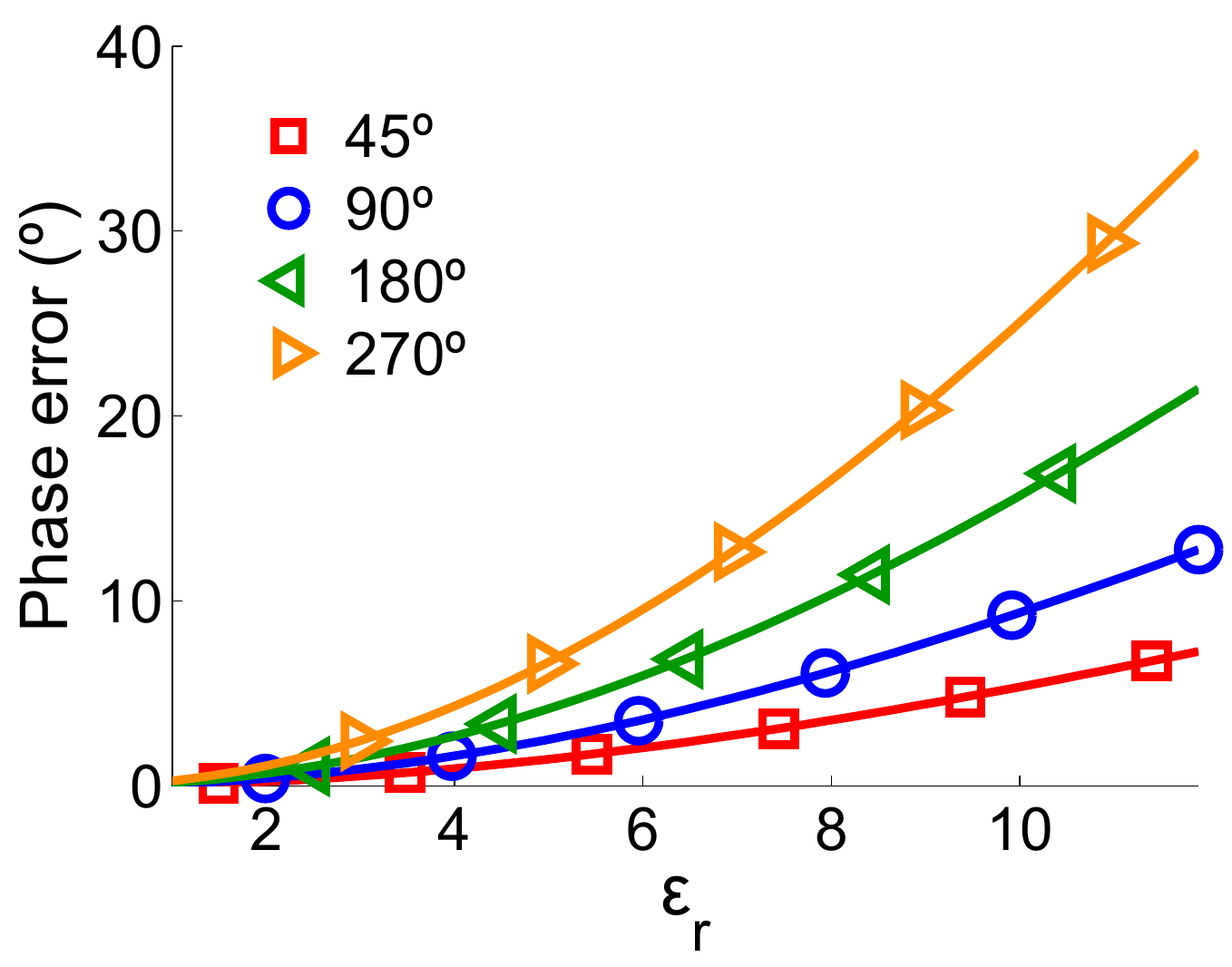}} \\
\subfloat[]{\label{fig: shifterd}
\includegraphics[width=0.6\columnwidth]{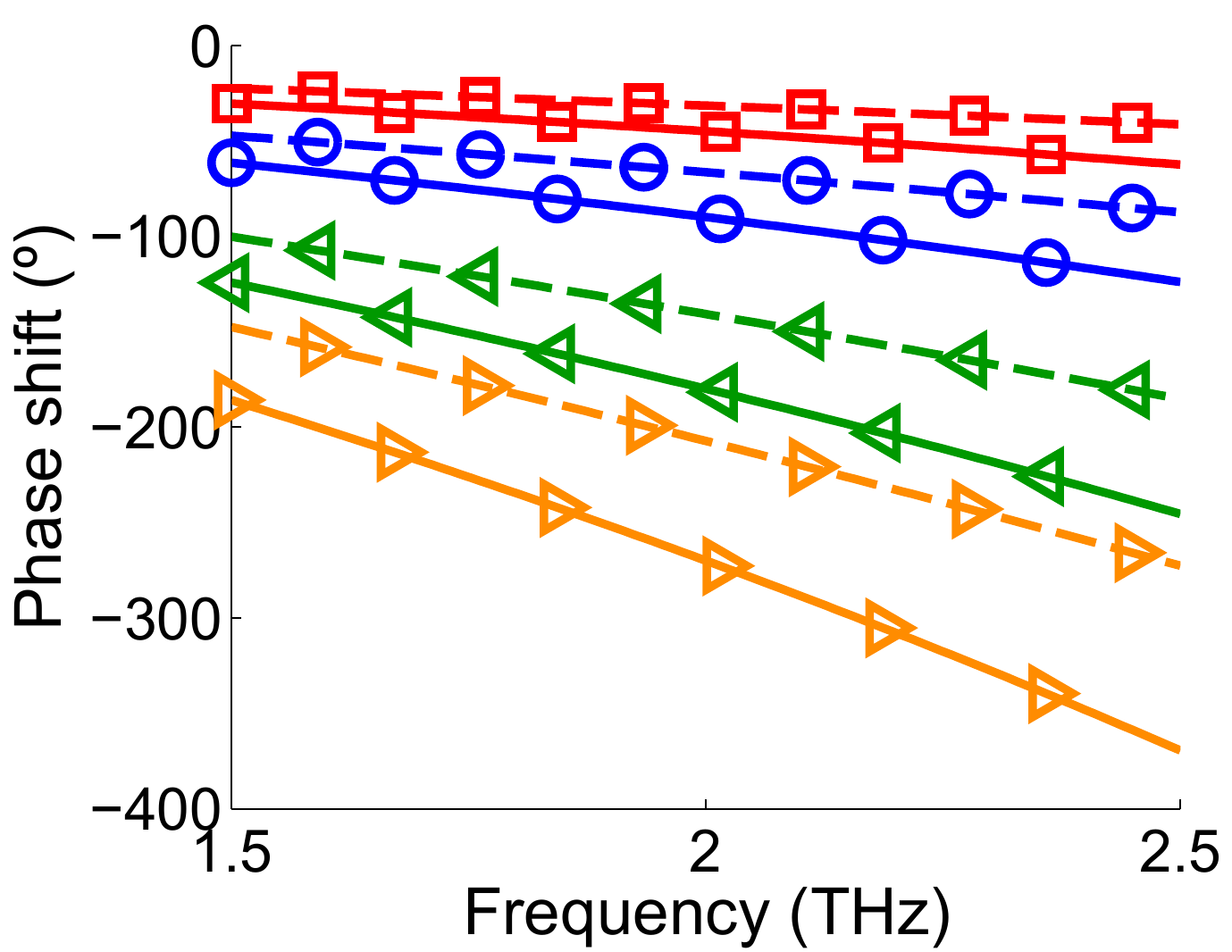}}
\subfloat[]{\label{fig: shiftere}
\includegraphics[width=0.6\columnwidth]{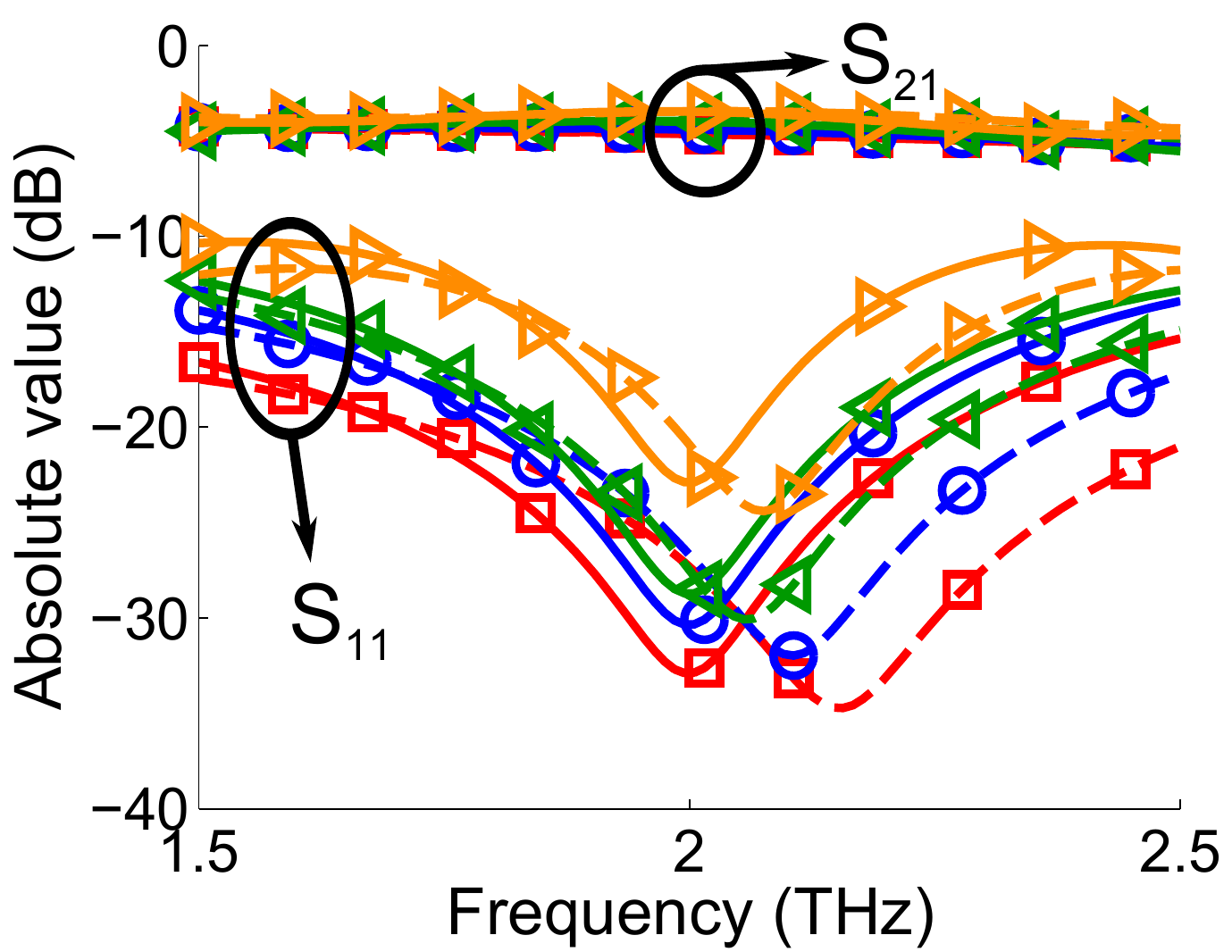}}
\subfloat[]{\label{fig: shifterf}
\includegraphics[width=0.6\columnwidth]{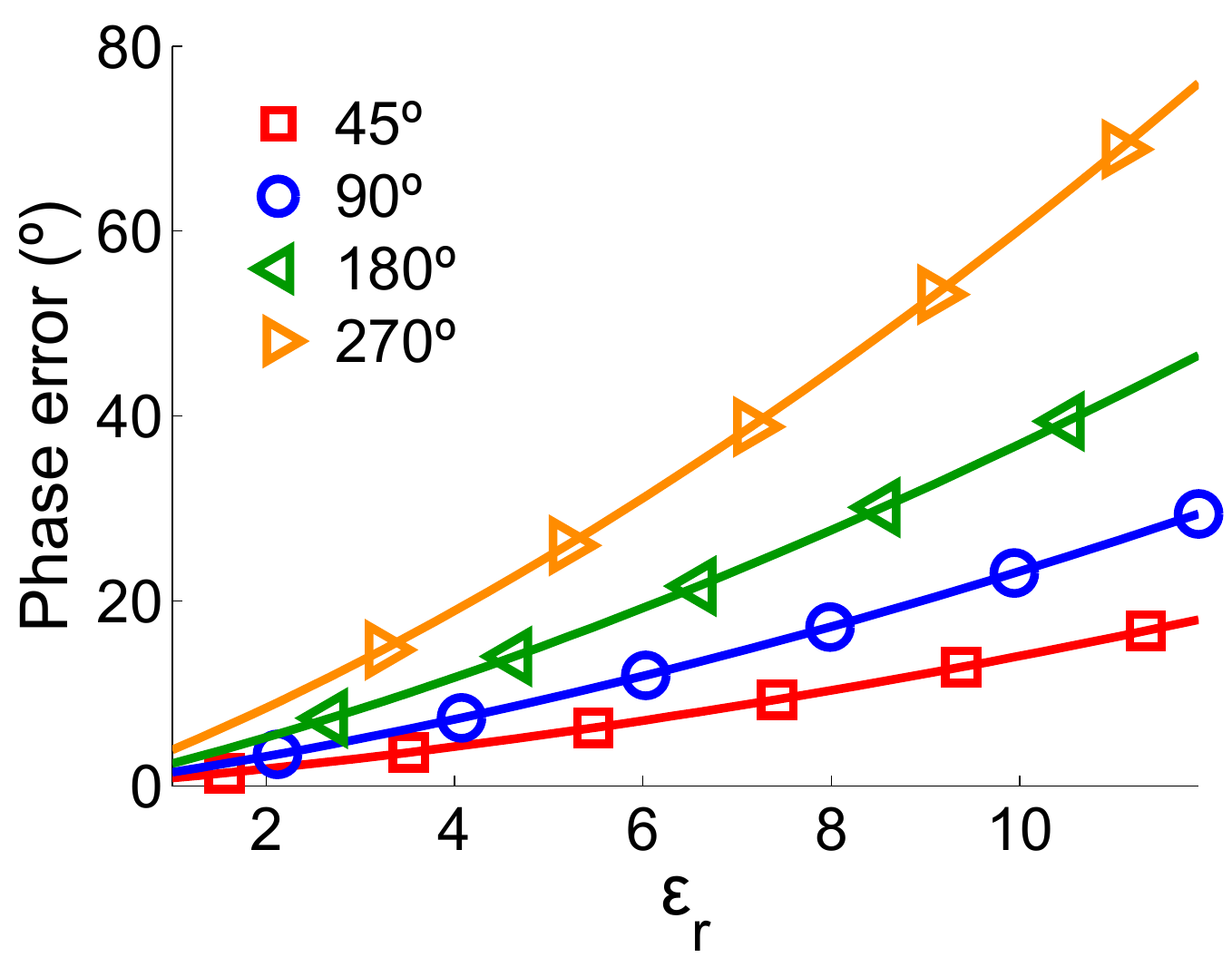}}
\caption{Influence of spatial dispersion in the response of the digital load-line phase shifters described in Table \ref{tab: shifter}. The upper row is related to the single sheet implementation: (a) phase difference between ports, (b) scattering parameters, and (c) phase error due to spatial dispersion versus the permittivity of the surrounding media. (d)-(f) show the same data for the PPW implementation. Parameters are $d = 100$~nm, $\tau = 1$~ps, $\varepsilon_{r1} = \varepsilon_{r2} = 11.9$ and $T=300$~K (solid line - results neglecting
spatial dispersion effects, dashed line - results including spatial dispersion effects).} \label{fig: shifter}
\end{figure*}
\begin{table}
\renewcommand{\arraystretch}{1.2}
\caption{Parameters of the 3-bit phase shifter ($\varepsilon_r = 11.9$).}
\label{tab: shifter}
\centering
\begin{tabular}{c c c c c}
\hline\hline
& $l_{sheet}(\mu m)$ & $ \mu_{c,sheet}$ (eV) & $l_{ppw}(\mu m)$ & $\mu_{c,ppw}$ (eV) \\
\hline
Ports & $0.25$ & $0.05$ & $0.25$ & $0.05$ \\
Bit 1 & $0.8$ & $0.067$ & $0.46$ & $0.08$ \\
Bit 2 & $0.95$ & $0.084$ & $0.55$ & $0.11$ \\
Bit 3 & $1.27$ & $0.11$ & $0.73$ & $0.19$ \\
\hline\hline
\end{tabular}
\end{table}
Let us consider a digital load-line graphene-based phase shifter where $N$ gated sections of the waveguide allow for $2^N$ relative phase shift states \cite{Chen13_phase_shifters}. Specifically, we implement a $3$-bit phase shifter with shifts of $45^\circ$, $90^\circ$ and $180^\circ$ at the operation frequency $f=2$~THz. The characteristics of each waveguide section,
in terms of chemical potential and pad length,
are obtained following the approach described in \cite{Chen13_phase_shifters}, and are shown in Table \ref{tab: shifter} considering a local graphene conductivity model. This design assumes that the host waveguides are embedded in Si ($\varepsilon_r=11.9$), which is a realistic situation. Note that in \cite{Chen13_phase_shifters} spatial dispersion was safely neglected because the structures were standing in free-space.
Figs.~\ref{fig: shiftera}-\ref{fig: shifterb} illustrate the performance of this phase shifter in the single layer implementation (Fig. \ref{fig:_MainFiga}). Ignoring spatial dispersion leads to large errors in the relative phase shifts between the ports, although no significant error occurs in the transmitted and reflected power. Fig \ref{fig: shifterc} illustrates the error in the phase shift at the operation frequency for various phase shifters designed with different surrounding permittivity. As can be observed, the influence of neglecting spatial dispersion significantly increases for higher permittivities values. Figs. \ref{fig: shifterd}-\ref{fig: shifterf} show the same information for the PPW implementation (Fig. \ref{fig:_MainFigb}). A larger error in the phase shift is observed in this case, which is due to the higher confinement of the quasi-TEM mode. Importantly, note that these errors can be easily compensated by simply using the correct spatially dispersive conductivity in the design procedure.
\subsection{\textit{\textsf{Graphene-Based Low-Pass Filters}}}
\begin{figure*}[!t] \centering
\subfloat[]{\label{fig: filtersa}
\includegraphics[width=0.6\columnwidth]{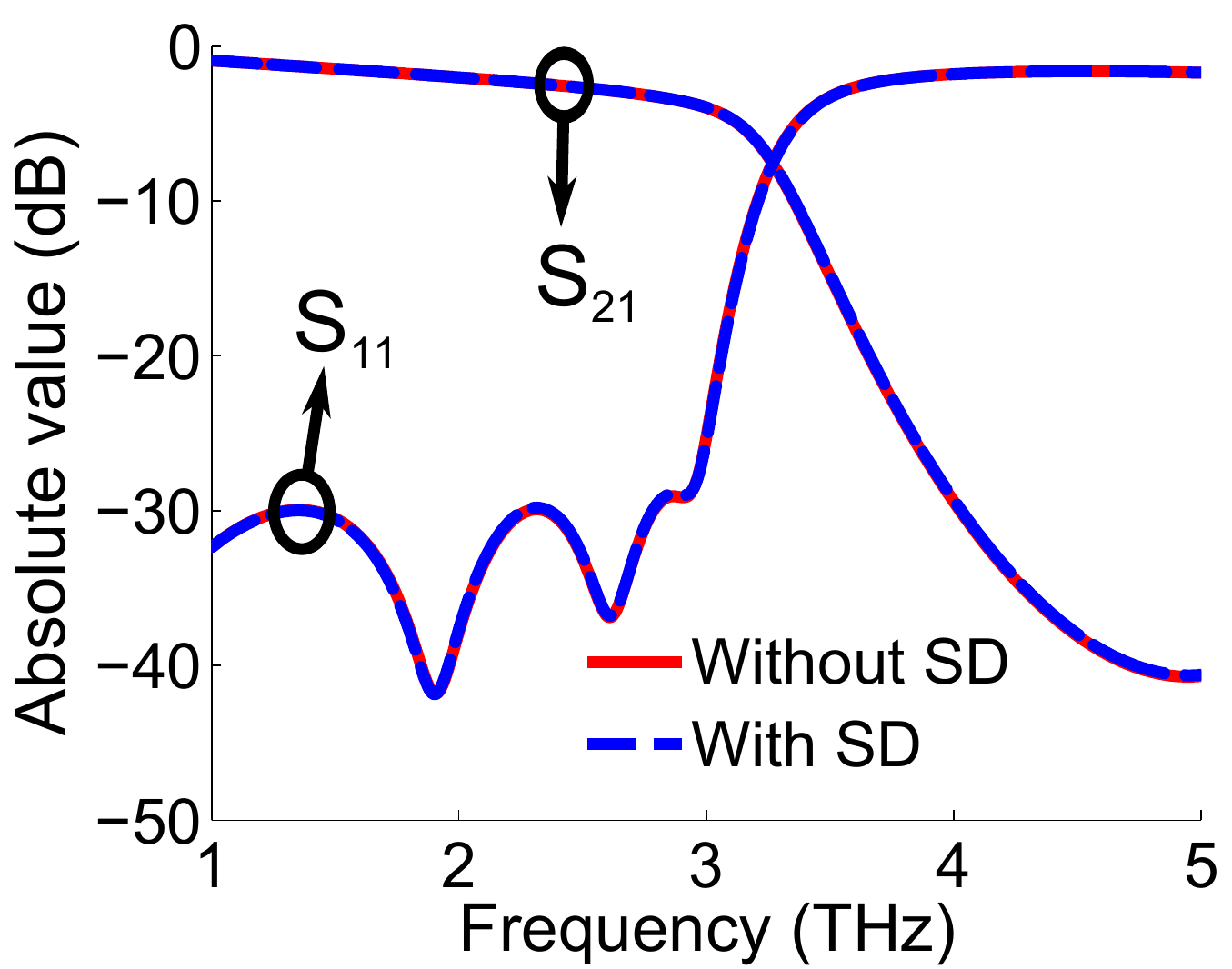}}
\subfloat[]{\label{fig: filtersb}
\includegraphics[width=0.6\columnwidth]{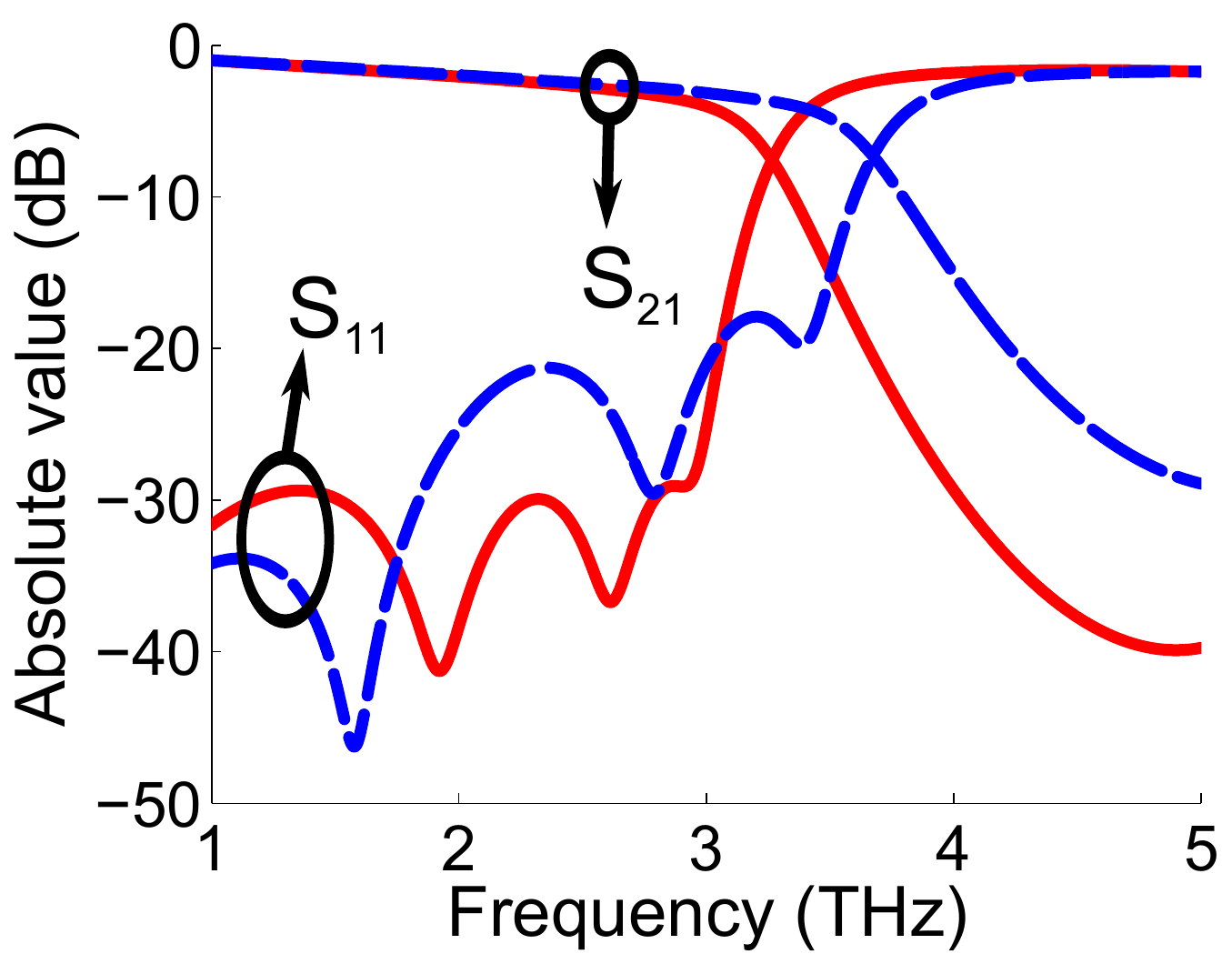}}
\subfloat[]{\label{fig: filtersc}
\includegraphics[width=0.6\columnwidth]{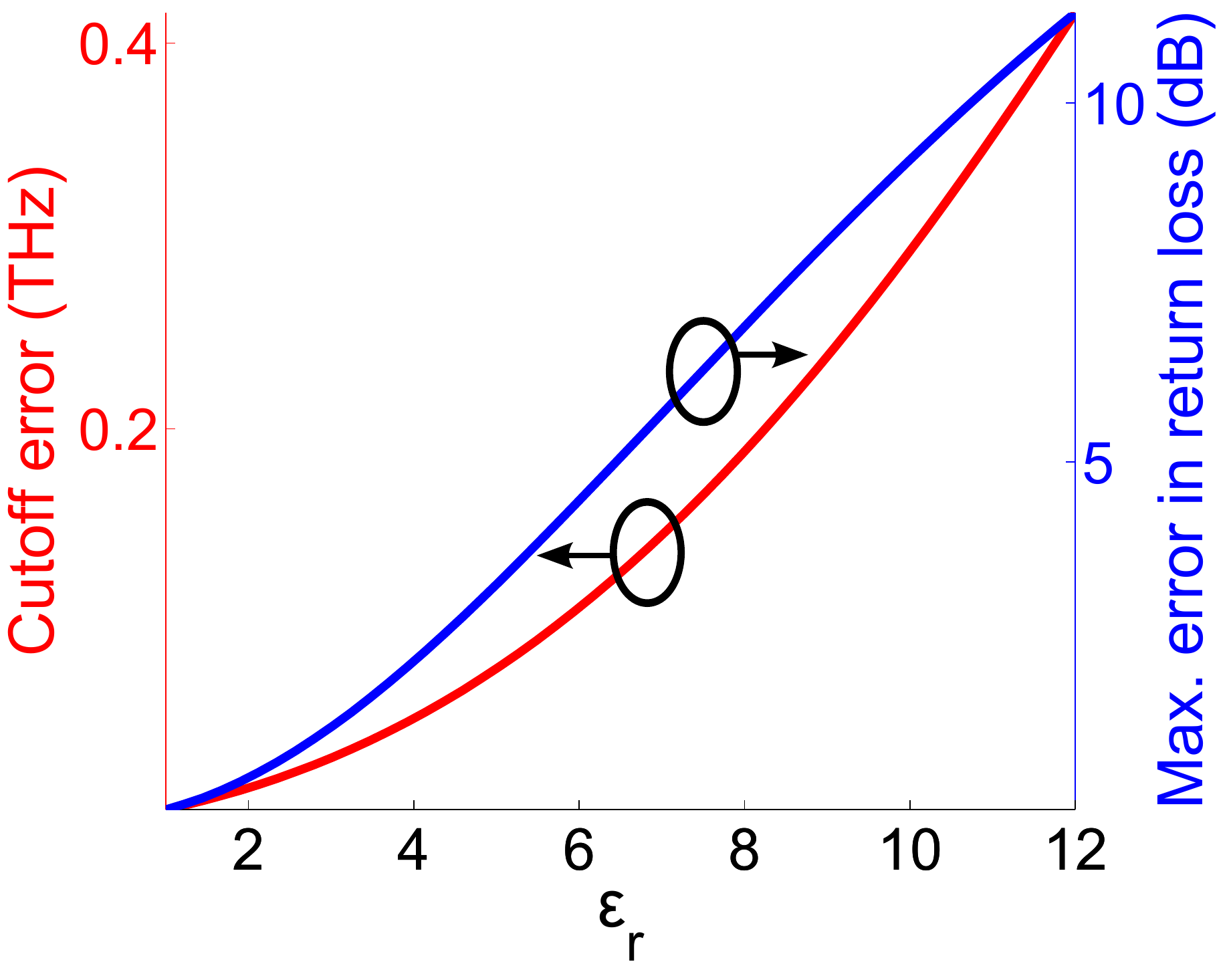}} \\
\subfloat[]{\label{fig: filtersd}
\includegraphics[width=0.6\columnwidth]{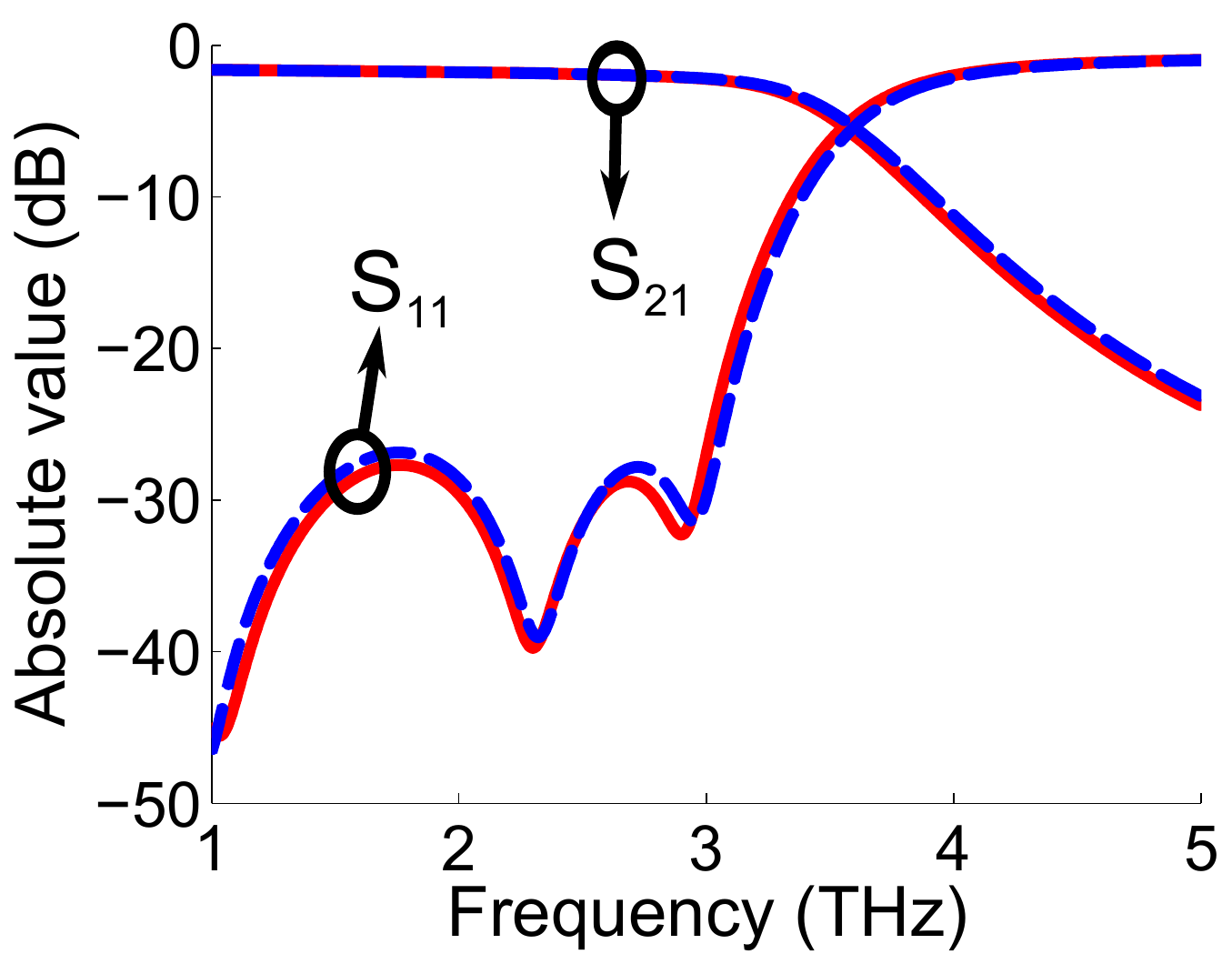}}
\subfloat[]{\label{fig: filterse}
\includegraphics[width=0.6\columnwidth]{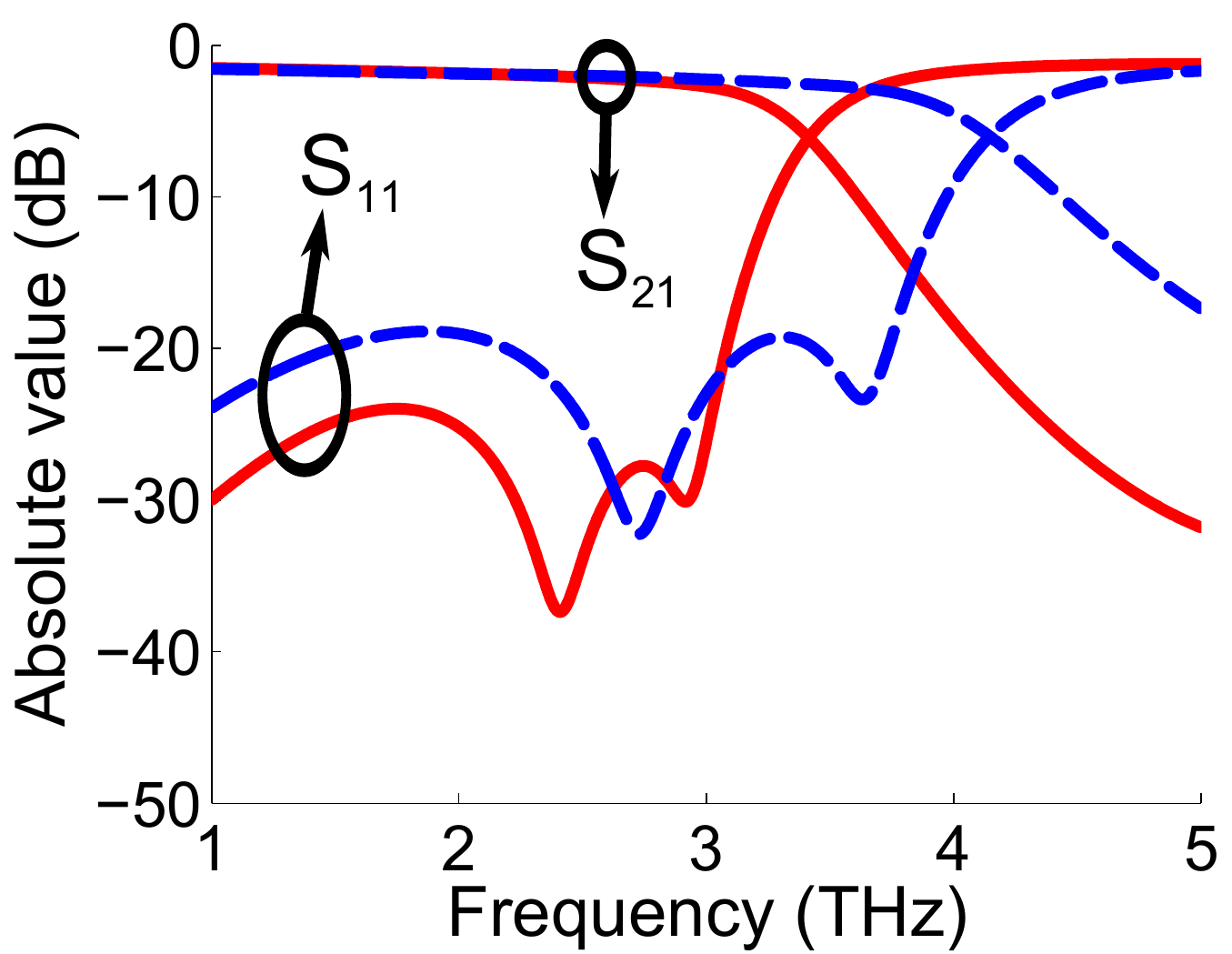}}
\subfloat[]{\label{fig: filtersf}
\includegraphics[width=0.6\columnwidth]{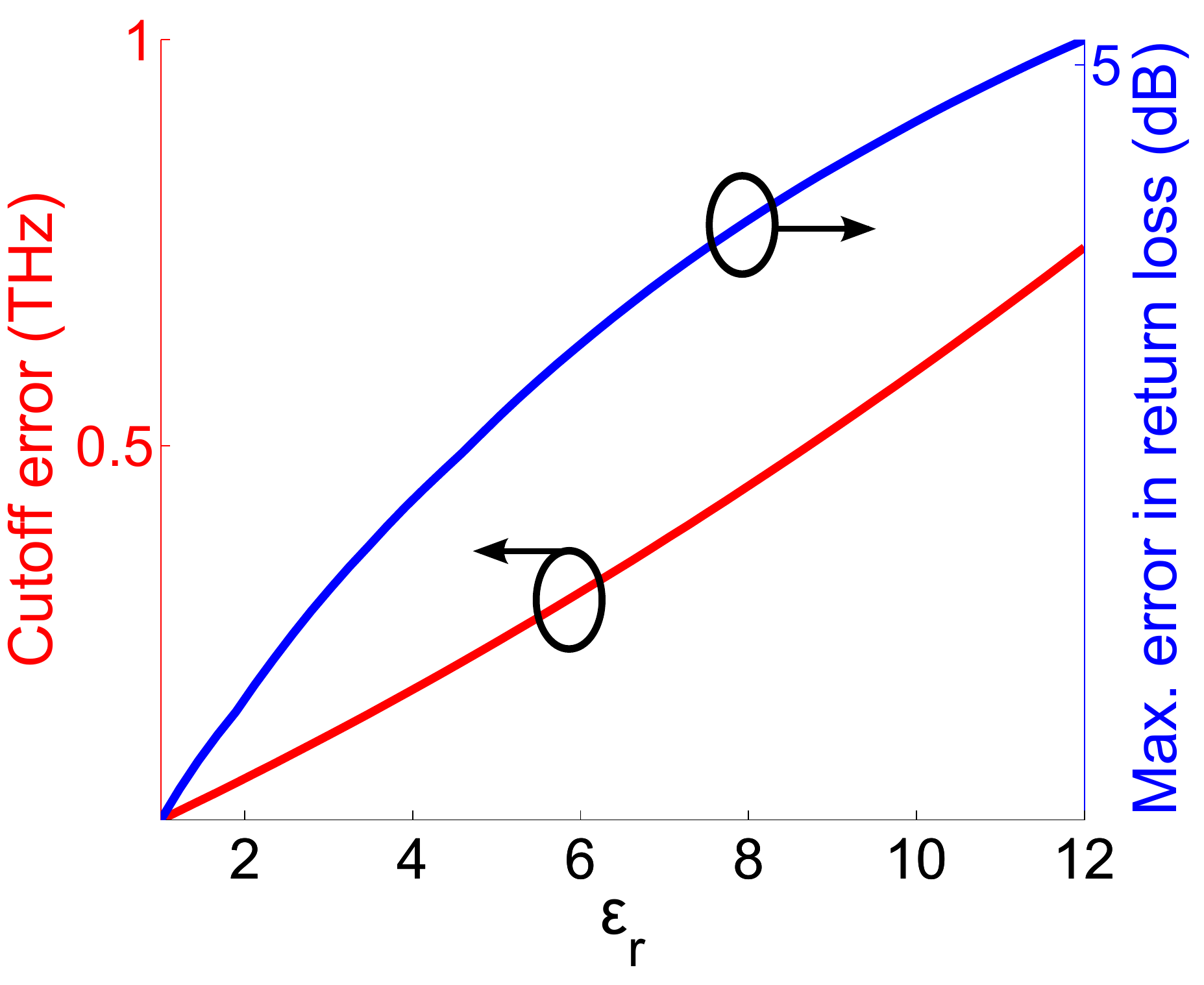}}
\caption{Influence of spatial dispersion in the response of $7^{th}$ degree graphene-based low-pass filters. Scattering parameters of the single sheet implementation in (a) free-space and (b) embedded in Si ($\varepsilon_{r1} = \varepsilon_{r2} = 11.9$, see Table~\ref{tab: filters 2}). (c) Error in the cutoff frequency and maximum in-band reflection due to spatial dispersion as a function of surrounding permittivity. (d)-(f) show the same data for the PPW implementation (see Table~\ref{tab: filters 2}). Parameters are $d = 100$~nm, $\tau = 1$~ps and $T=300$~K (solid line - results neglecting
spatial dispersion effects, dashed line - results including spatial dispersion effects).}\label{fig: filters}
\end{figure*}

Let us consider graphene-based low-pass filters in the THz, as proposed in \cite{Diego13}. Specifically, we implement a $7^{th}$ degree filter with cutoff frequency $f_c=3$~THz. For the sake of comparison, we design this filter in free-space ($\varepsilon_r=1$) and embedded in Si ($\varepsilon_r=11.9$).
The design parameters of the Si-embedded filter, computed following \cite{Diego13}, are shown in Table \ref{tab: filters 2}.
Figs.~\ref{fig: filtersa}-\ref{fig: filtersb} show the frequency response corresponding to the single graphene sheet filters, standing in free-space and embedded in Si, respectively. As expected, for $\varepsilon_r = 1$ spatial dispersion proves to be irrelevant. On the other hand, for $\varepsilon_r = 11.9$ the filter response severely deteriorates, up-shifting its cutoff frequency and unevenly
increasing the reflection
\begin{table}[]
\renewcommand{\arraystretch}{1.2}
\caption{Parameters of the $7^{th}$ degree filter ($\varepsilon_r = 11.9$).}
\label{tab: filters 2}
\centering
\begin{tabular}{c c c c c}
\hline\hline
& $l_{sheet}(nm)$ & $ \mu_{c,sheet}$ (eV) & $l_{ppw}(nm)$ & $\mu_{c,ppw}$ (eV) \\
\hline
Ports & $50$ & $0.13$ & $50$ & $0.14$ \\
1,7 & $81$ & $0.08$ & $66$ & $0.09$ \\
2,6 & $247$ & $0.26$ & $142$ & $0.32$ \\
3,5 & $47$ & $0.04$ & $43$ & $0.03$ \\
4 & $308$ & $0.33$ & $175$ & $0.46$ \\
\hline\hline
\end{tabular}
\end{table}

throughout the passband. Note that the presence of spatial dispersion prevents the total compensation of this latter effect, due to the higher non-linear nature of the mode's propagation constant and characteristics impedance. Fig \ref{fig: filtersc} illustrates the error in the filter's cutoff frequency and the maximum return loss in the filter's passband for various low-pass filters designed using dielectrics with increasing permittivity values. Similarly to the case of the phase-shifters, the influence of spatial dispersion increases when the permittivity of the surrounding medium increases. Figs. \ref{fig: filtersd}-\ref{fig: filtersf} show the same study for the PPW implementation. As expected, a larger shift in the cutoff frequency is observed in this case, but interestingly, the maximum error of the return loss within the passband is lower. Contrary to the single sheet implementation, we have verified that a uniform level of in-band return loss can be achieved using graphene-based PPW, because the characteristic impedance of each spatially-dispersive transmission line section remains
more

linear with frequency.
This indicates
that the use of graphene PPW structures could be advantageous
over the use of single sheet structures, when low return losses
are essential in lowpass filter applications.


\section{\textit{\textsf{\textbf{Conclusion}}}}
We have studied the influence of non-local effects in the response and performance of plasmonic graphene THz devices. Following previous works, we have focused on graphene-based phase shifters and low pass filters, necessary elements for THz communication and sensing systems. Due to the extremely slow waves supported by graphene-based waveguides in the presence of high permittivity media, spatial dispersion becomes a significant mechanism of propagation that modifies the expected behavior of these devices by up-shifting their operation frequency, limiting their tunable range, and degrading their frequency response. Consequently, spatial dispersion must be accurately taken into account in the development of graphene-based plasmonic THz devices.
\section*{\textit{\textsf{\textbf{Acknowledgments}}}}
This work was supported by the EU FP7 Marie-Curie IEF grant Marconi, with ref. $300966$, Spanish Ministry of Education under grant TEC2010-21520-C04-04, and European Feder Fundings.

\end{document}